\documentclass[
    ,final            
  ]
  {aipproc}

\layoutstyle{6x9}

\begin{document}

\title[CoRoT-Exo-2a activity]{CoRoT and stellar activity: preliminary results from the modelling of CoRoT-Exo-2a}

\classification{97.20.Jg -- 97.10.Jb -- 97.10.Kc -- 97.10.Qh -- 97.82.Fs -- 97.10.Ld}
\keywords      {Main-sequence: late-type stars -- stellar activity -- stellar rotation -- surface features -- planets -- magnetic fields}

\author{A.~F.~Lanza}{
  address={INAF-Osservatorio Astrofisico di Catania, Via S.~Sofia, 78, 95123 Catania, Italy}
}

\author{I.~Pagano}{
  address={INAF-Osservatorio Astrofisico di Catania, Via S.~Sofia, 78, 95123 Catania, Italy}
}

\author{G.~Leto}{
  address={INAF-Osservatorio Astrofisico di Catania, Via S.~Sofia, 78, 95123 Catania, Italy}
}

\author{S.~Messina}{
  address={INAF-Osservatorio Astrofisico di Catania, Via S.~Sofia, 78, 95123 Catania, Italy}
}

\author{P.~Barge}{
  address={Laboratoire d'Astrophysique de Marseille, UMR~6110, CNRS, Universit\'e de Provence, Traverse du Siphon, 13376 Marseille, France}
}

\author{A.~Baglin}{
  address={LESIA, CNRS UMR~8109, Observatoire de Paris, 5 place J.~Janssen, 92195 Meudon, France} 
}

\begin{abstract}
We present a preliminary analysis of the photospheric activity of CoRoT-Exo-2a, a young G7V star accompanied by a transiting hot Jupiter recently discovered by CoRoT.  We apply spot modelling techniques developed for the analysis of the Sun as a star and  capable to extract from CoRoT high precision light curves information on the variation of  the total spotted area and the longitude of active regions along the 142 days of the observations.  This preliminary analysis shows that the active regions form within two active longitudes separated  by about $180^{\circ}$ and rotating with periods of 4.5221 and 4.5543 days, respectively,  and that the total spotted area oscillates with a period  of about 28.9 days. 
\end{abstract}

\maketitle


\section{Introduction}
CoRoT is a space experiment devoted to asteroseismology and extrasolar planet search through the observations of planetary transits. It has recently discovered a hot Jupiter, CoRoT-Exo-2b,  orbiting with a period of 1.743 days around a main-sequence G7 star which displays a remarkable photospheric activity \citep{Alonsoetal08,Bouchyetal08}. Given the late spectral type and short rotation period (about 4.5 days) of the star, its activity is regarded as the manifestation of magnetic fields in the atmosphere, amplified and modulated by a hydromagnetic dynamo. In this paper, we present some preliminary results about the spot modelling of such a star, indicated as CoRoT-Exo-2a, which is a good proxy for the young Sun, probably at an age of approximately 0.5 Gyr \citep{Bouchyetal08} . A detailed account of the results obtained from the spot modelling of the light curve of CoRoT-Exo-2a will be provided in \citep{Lanzaetal08}.  
  
\section{Observations}
CoRoT-Exo-2a was observed from May 16 to October 5, 2007. We extracted from the data archive the N2 chromatic light curves having a sampling of 512 s during the first week and 32 s thereinafter. The  red, green and blue fluxes were summed up to get the white light flux and transits were removed by means of the ephemeris and parameters of \citep{Alonsoetal08}. We initially disregarded all data points at a distance from the mean greater than 4.2 standard deviations of the whole data set; then, we subtracted a moving-median filtered version of the light curve (box-car extension: 1 orbital period of the satellite, i.e., 6184 s) and discarded the points at a distance greater than 3 standard deviations of the residuals. Finally,  we computed normal points by binning the data on a time interval of one orbital period of the satellite. 

\section{Spot modelling}

We apply the maximum entropy (hereinafter ME) spot modelling method of \citep{Lanzaetal07}, to  whom we refer the reader for more details. The model assumes that the photosphere of the star is subdivided into 200 surface elements of size $18^{\circ} \times 18^{\circ}$ which are covered by cool spots, solar-like faculae and unperturbed photosphere. Following \citep{Gondoin08} and 
\citep{Lockwoodetal07}, who suggested that faculae have a secondary role in the light variations of late-type stars significantly more active than the Sun, we neglect solar-like faculae in this preliminary study and defer their  consideration to \citep{Lanzaetal08}.  The fraction of the area of each surface element covered by cool spots is given by the filling factor $f$,  so $1-f$ is the fraction occupied by the unperturbed photosphere. A stable and unique map, specified by the vector of the filling factor values ${\vec f}$, is derived by minimizing a linear combination of the chi square $\chi^{2}$ and the entropy functional $S$ , i.e.: 
\begin{equation}
Q = \chi^{2} ({\vec f}) - \lambda S ({\vec f}), 
\label{object_funct}
\end{equation} 
where the Lagrangian multiplier $\lambda > 0$ rules the trade-off between  light curve fitting, as  measured by the $\chi^{2}$, and the regularization, as measured by the entropy functional 
$S$.  The optimal Lagrangian multiplier is determined iteratively by making the mean of the residuals deviate by  one standard error of the normal points from the value obtained without regularization \citep[see ][]{Lanzaetal08}.  
    The stellar parameters are taken from \citep{Alonsoetal08} and \citep{Bouchyetal08}. The stellar rotation axis is assumed to be perpendicular to the orbital plane of the planet. The spot temperature is assumed  $\sim 540$ K below that of the unperturbed photosphere. The light curve is divided into 45 subsets of duration 3.15611 days because the rapid change of the spot pattern does not allow us to obtain a good fit with longer time intervals \citep[cf. ][for the case of the Sun]{Lanzaetal03,Lanzaetal04}. The Lomb-Scargle periodogram gives a period of the rotational  modulation of $4.52 \pm 0.14$ days.

\section{Results}
\begin{figure}
  \includegraphics[height=.5\textheight,angle=90]{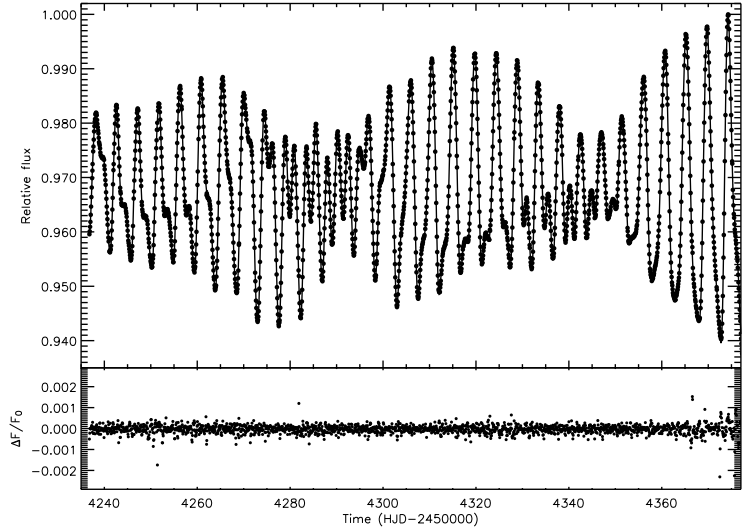}
  \caption{
Upper panel: observations of CoRoT-Exo-2a (filled dots) versus time with the best fit obtained with our ME spot model (solid line); the flux is normalized to the maximum value observed along the 142-d time series; lower panel: the residuals of the best fit in relative flux units. 
}
\label{Lanzaf1}
\end{figure}

\begin{figure}
  \includegraphics[height=.5\textheight,angle=90]{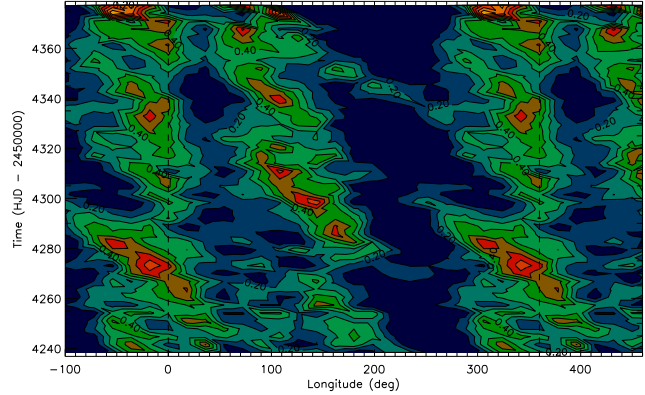}
  \caption{
Isocontours of the ratio $f/f_{\rm max}$, where $f$ is the spot covering factor and $f_{\rm max} = 0.0163$ its maximum value, versus time and longitude for our ME spot models. The two dashed vertical lines mark longitudes $0^{\circ}$ and 360$^{\circ}$ beyond which the distributions are repeated to easily identify spot migration. The contour levels are separated by 10 percent of the maximum filling factor, with light yellow indicating the maximum covering factor and dark blue the minimum.   
}
\label{Lanzaf2}
\end{figure}

\begin{figure}
  \includegraphics[width=.33\textwidth,height=.3\textheight,angle=90]{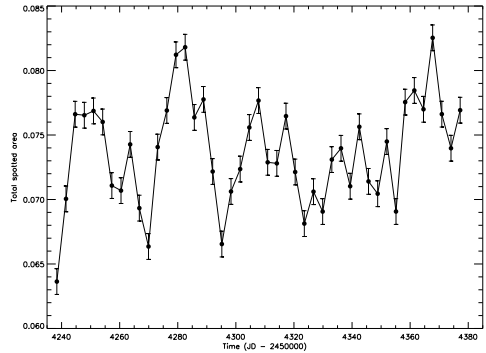}
  \caption{
The variation of the total spotted area versus time for the ME spot models. The area is measured in units of the stellar photosphere. The error bars account only for random errors in the area and have a semiamplitude of 3 standard deviations.  
}
\label{Lanzaf3}
\end{figure}

The sequence of best fits obtained with our ME model is shown in Fig.~\ref{Lanzaf1} together with the residuals versus time. The best fit is always very good, with an average standard deviation of $2.26 \times 10^{-4}$  in relative units. Since the inclination of the stellar rotation axis is very close to $90^{\circ}$, only the distribution of the spotted area vs. longitude can be derived through the spot modelling. We plot the normalized spot filling factor versus longitude and time in Fig.~\ref{Lanzaf2}. The longitude increases in the same direction of the stellar rotation and the orbital motion of the planet. The adopted rotation period for the model star is 4.5221 days. 
The star shows two active longitudes one of  which does not migrate appreciably in the adopted reference frame, i.e., has a rotation period of 4.5221 days, while the other shows a slow migration indicating a rotation period of 4.5543 days. Interpreted in terms of surface differential rotation, this indicates a significantly smaller relative amplitude than in the Sun, i.e., about 1 percent. Individual spots show an angular velocity about 1.3 percent smaller than that of the active longitudes. 
The total spotted area is plotted vs. time in Fig.~\ref{Lanzaf3} and shows a cyclic oscillation with a period of $28.9 \pm 4.8$ days, as derived from Lomb-Scargle periodogram. It is interesting to note that such a period is close to 10 times the synodic period of the planet as seen by the active longitude pattern rotating in 4.5221 days. This may suggest a possible star-planet magnetic interaction (see \citep{Lanza08} and \citep{Lanzaetal08} for a possible interpretation).   
 It is important to notice that a different spot temperature gives different absolute values of the spotted area, but does not affect the cyclic variation we have found. Such a variation is not readily apparent from the light modulation because two spots on opposite hemispheres are usually responsible for the flux variations observed in CoRoT-Exo-2a, thus the light curve amplitude is not a good indicator of the total spotted area in this star. 


\begin{theacknowledgments}
The present study is based on observations obtained with CoRoT, a space project de\-ve\-lo\-ped and operated by the French Space Agency, CNES, with partecipation of the Science Program of ESA, ESTEC/RSSD, Austria, Belgium, Brazil, Germany and Spain. AFL, IP, GL and SM have been partially supported by the Italian Space Agency (ASI) under contract ASI/INAF I/015/07/0, work package 3170. 
\end{theacknowledgments}

\end{document}